\documentclass[final,12pt]{elsarticle}

\usepackage{amssymb}
\usepackage{amsthm}

\usepackage[mathlines]{lineno} 
\usepackage{amsmath}           

\usepackage{etoolbox} 

\newcommand*\linenomathpatch[1]{%
  \cspreto{#1}{\linenomath}%
  \cspreto{#1*}{\linenomath}%
  \csappto{end#1}{\endlinenomath}%
  \csappto{end#1*}{\endlinenomath}%
}
\newcommand*\linenomathpatchAMS[1]{%
  \cspreto{#1}{\linenomathAMS}%
  \cspreto{#1*}{\linenomathAMS}%
  \csappto{end#1}{\endlinenomath}%
  \csappto{end#1*}{\endlinenomath}%
}

\expandafter\ifx\linenomath\linenomathWithnumbers
  \let\linenomathAMS\linenomathWithnumbers
  \patchcmd\linenomathAMS{\advance\postdisplaypenalty\linenopenalty}{}{}{}
\else
  \let\linenomathAMS\linenomathNonumbers
\fi

\linenomathpatch{equation}
\linenomathpatchAMS{gather}
\linenomathpatchAMS{multline}
\linenomathpatchAMS{align}
\linenomathpatchAMS{alignat}
\linenomathpatchAMS{flalign}

\usepackage{todonotes}
\usepackage{graphicx} 
\usepackage{verbatim}
\usepackage{kbordermatrix}
\usepackage{tikz}
\usepackage{hyperref}

\newtheorem{lemma}{Lemma}

\newtheorem{thm}{Theorem}
\newtheorem{prop}{Proposition}
\newtheorem{cor}{Corollary}
\theoremstyle{definition}
\newtheorem{definition}{Definition}

\newcommand{\opt}{\operatorname{OPT}}

\journal{Operations Research Letters}

\begin{document}

\begin{frontmatter}



\title{Unconstrained Traveling Tournament Problem is APX-complete}


\author[inst1]{Salomon Bendayan}

\affiliation[inst1]{
	    email={s2benday@uwaterloo.ca},
	    organization={Department of Combinatorics and Optimization, University
            of Waterloo},
            addressline={200 University Avenue West}, 
            city={Waterloo},
            postcode={N2L 3G1}, 
            state={Ontario},
            country={Canada}
	    }

\author[inst2]{Joseph Cheriyan}
\author[inst3]{Kevin K. H. Cheung}

\affiliation[inst2]{
	    email={jcheriyan@uwaterloo.ca},
	    organization={Department of Combinatorics and Optimization, University
            of Waterloo},
            addressline={200 University Avenue West}, 
            city={Waterloo},
            postcode={N2L 3G1}, 
            state={Ontario},
            country={Canada}
	    }

\affiliation[inst3]{
	    email={kcheung@math.carleton.ca},
	    organization={School of Mathematics and Statistics, Carleton
            University},
            addressline={1125 Colonel By Drive}, 
            city={Ottawa},
            postcode={K1S 5B6}, 
            state={Ontario},
            country={Canada}
	    }

\begin{abstract}
The unconstrained Traveling Tournament Problem is APX-complete.
\end{abstract}


\begin{keyword}
Traveling Tournament Problem \sep APX-complete \sep Approximation algorithms
     \sep Traveling Salesman Problem
\MSC 90C27 \sep 68Q25
\end{keyword}

\end{frontmatter}


\section{Introduction}
\label{sec:sample1}

Easton, Nemhauser and Trick~\cite{ttp} introduced the Traveling
Tournament Problem (TTP) in 2001.  
In fact, there are several versions of the TTP;
the version formulated in \cite{ttp} is called TTP($3$), and another
(simpler and fundamental) version is the unconstrained TTP, denoted UTTP.
Over the years, many papers have been published on the TTP; the
majority of the papers are on computational and modeling issues,
while a few papers address issues pertaining to the computational complexity of TTP.

Although it is natural to conjecture that the
TTP is at least as hard as the well-known (metric) Traveling Salesman
Problem (TSP), rigorous proofs were published many years after the
appearance of \cite{ttp}.
Bhattacharyya~\cite{uttpnp} first showed that
UTTP is NP-hard (he published a technical report in 2009).
Subsequently, Thielen and Westphal \cite{ttpnp} proved that TTP($3$)
is NP-hard, via a reduction from 3-SAT.  Recently, Chatterjee~\cite{npttpk}
has reported that TTP($k$) is NP-hard for all fixed $k>3$, using a
reduction from $k$-SAT.

In this paper, our goal is to show that UTTP is APX-complete, thus ruling out
a PTAS (polynomial-time approximation scheme) for solving UTTP though a
2.75-approximation algorithm for UTTP does exist~\cite{uttp275}.
The (1,2)-TSP is the special case of the (metric) TSP where every
inter-city distance is one or two.
Papadimitriou and Yannakakis~\cite{PY93} showed that (1,2)-TSP is
APX-complete, via the notion of an L-reduction.  We start with this
result of \cite{PY93}, and we construct an L-reduction from (1,2)-TSP to UTTP.

Our construction is inspired by Bhattacharyya's construction for
proving that UTTP is NP-hard.
In fact, our construction may be viewed as a version of Bhattacharyya's
construction and described in a more verbal way.
We refer to Bhattacharyya~\cite{uttpnp} for key claims and their proofs.

The results of this paper appear in preliminary form in \cite{salomon}.

\subsection{The Traveling Tournament Problem}

Given $n$ teams, with $n$ being an even number, a
  \textit{double round-robin tournament} is a schedule of games
played between pairs of teams such that each pair of teams plays two
games.  Clearly, each team plays $2(n-1)$ games, so the schedule has
$n\times(n-1)$ games.  The schedule has $2(n-1)$ days, where each day
has $n/2$ games; thus each team plays on each day.
For each game, one of the teams is designated the \textit{home} team
and the game is played at the city of the home team;
and the other team is designated the \textit{away} team;

For each team, a series of consecutive away games is called a
\textit{road trip}, while a series of consecutive home games is
called a \textit{home stand}.
(Note that the length of a road trip (or home stand) is given by
the number of opponents played and not by the distance travelled
by the team.)

For all  pairs of teams $i$ and $j$, the distance between the city
of $i$ and the city of $j$ is given in an $n \times n$ matrix $D$;
the inter-city distances are assumed to satisfy the triangle inequality.

When playing an away game, a team travels from its home city to the
opponent's city. When a team is playing consecutive away games, it
travels directly from the city of one opponent to the city of the
next opponent. At the conclusion of its road trip, the team returns
to its home city.

The \textit{Traveling Tournament Problem (TTP)} can be succinctly described as follows:

\textbf{Input: } An even number of teams, $n$; an $n \times n$ integral
distance matrix, $D$; integer parameters $L,U$.

\textbf{Output: } A double round-robin tournament of the $n$ teams such that:

\begin{itemize}
    \item every team plays every other team once at its home city
    and once at the city of the opponent

    \item the length of every home stand and the length of every road trip
	is between $L$ and $U$, and

    \item the total distance travelled by the teams is minimized. 
	(At the start of the tournament before the first day, each team
	is assumed to be at its home city, and at the end of tournament
	after the $2(n-1)$-th day, each team is assumed to return to its
	home city.)

\end{itemize}

A schedule satisfying the first constraint is a feasible double
round-robin tournament. The second constraint is called the
\textit{at-most} constraint when $L = 1$. The final constraint corresponds to
the objective function. There may be an additional constraint placed
on the tournament:

\begin{itemize}
    \item[-] \textit{No repeaters:}
    in the tournament schedule, there is no pair of teams $i$ and $j$ such that
	the two games between $i$ and $j$ are played on consecutive days.
\end{itemize}

The TTP was originally introduced with $L=1$ and $U = 3$.

In this paper, we focus on the variant of TTP with $L=1$ and $U=n-1$
called \textit{unconstrained TTP (UTTP)}.
Our main result holds whether or not the no-repeaters constraint is present.

\subsection{L-reductions and APX-completeness}

This subsection discusses the notions of L-reduction and APX-completeness.

\begin{definition}[cf. Definition 16.4 in \cite{approxalg}]
Given two NP optimization problems $\Pi$ and $\Pi'$, we have an
  \textit{L-reduction} from $\Pi$ to $\Pi'$ if for some parameters $a,b>0$:
\begin{enumerate}
    \item For each instance $I$ of $\Pi$ we can compute in polynomial time an instance $I'$ of $\Pi'$.
    \item $\opt(I'$) $\leq$ $a \cdot \opt(I$).
    \item Given a solution of value $V'$ to $I'$, we can compute in polynomial time a solution of value $V$ to $I$ such that
    \begin{equation*}
        |\opt(I) - V| \leq b|\opt(I') - V'|
    \end{equation*}
\end{enumerate}
\end{definition}

APX is the class of NP Optimization problems that have $O(1)$-approximation 
algorithms
(see Definition~3.9, Chapter~3.1 \cite{ACGKMP}).
A problem $\pi$ in APX is called APX-complete if there exists an
L-reduction from every problem in APX to $\pi$ (see Chapter~8.4 \cite{ACGKMP}).

\begin{prop}[Papadimitriou and Yannakakis~\cite{PY93}]\label{prop:PY}
(1,2)-TSP is APX-complete.
\end{prop}

\section{Construction of a UTTP instance from a (1,2)-TSP instance}
\label{construction}
Let $G$ be a complete graph with edge-costs in $\{1,2\}$.
Observe that the edge-costs form a metric; i.e they satisfy the triangle
inequality.

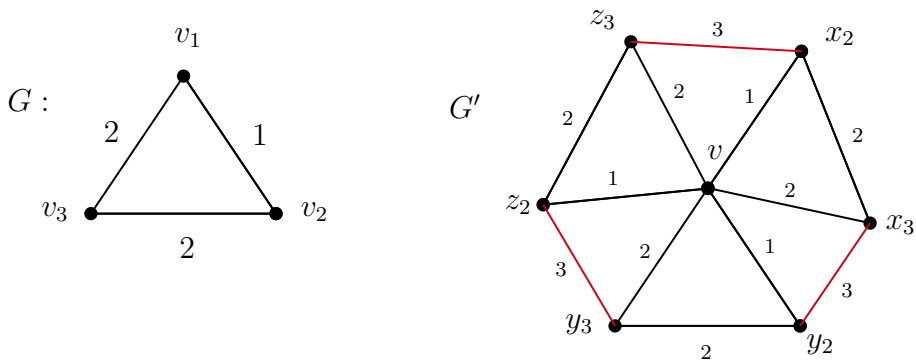
\begin{figure}[h] 
    \centering
\tikzset{every picture/.style={line width=0.75pt}} 

\begin{tikzpicture}[x=0.75pt,y=0.75pt,yscale=-0.85,xscale=0.85]

\draw   (139.17,59.33) -- (194,141) -- (84.17,141) -- cycle ;
\draw    (139.17,59.33) -- (194,141) ;
\draw [shift={(194,141)}, rotate = 56.12] [color={rgb, 255:red, 0; green, 0; blue, 0 }  ][fill={rgb, 255:red, 0; green, 0; blue, 0 }  ][line width=0.75]      (0, 0) circle [x radius= 3.35, y radius= 3.35]   ;
\draw [shift={(139.17,59.33)}, rotate = 56.12] [color={rgb, 255:red, 0; green, 0; blue, 0 }  ][fill={rgb, 255:red, 0; green, 0; blue, 0 }  ][line width=0.75]      (0, 0) circle [x radius= 3.35, y radius= 3.35]   ;
\draw    (194,141) -- (84.17,141) ;
\draw [shift={(84.17,141)}, rotate = 180] [color={rgb, 255:red, 0; green, 0; blue, 0 }  ][fill={rgb, 255:red, 0; green, 0; blue, 0 }  ][line width=0.75]      (0, 0) circle [x radius= 3.35, y radius= 3.35]   ;
\draw [shift={(194,141)}, rotate = 180] [color={rgb, 255:red, 0; green, 0; blue, 0 }  ][fill={rgb, 255:red, 0; green, 0; blue, 0 }  ][line width=0.75]      (0, 0) circle [x radius= 3.35, y radius= 3.35]   ;
\draw   (450.17,126) -- (505,207.67) -- (395.17,207.67) -- cycle ;
\draw    (450.17,126) -- (505,207.67) ;
\draw [shift={(505,207.67)}, rotate = 56.12] [color={rgb, 255:red, 0; green, 0; blue, 0 }  ][fill={rgb, 255:red, 0; green, 0; blue, 0 }  ][line width=0.75]      (0, 0) circle [x radius= 3.35, y radius= 3.35]   ;
\draw [shift={(450.17,126)}, rotate = 56.12] [color={rgb, 255:red, 0; green, 0; blue, 0 }  ][fill={rgb, 255:red, 0; green, 0; blue, 0 }  ][line width=0.75]      (0, 0) circle [x radius= 3.35, y radius= 3.35]   ;
\draw    (505,207.67) -- (421.17,207.67) -- (395.17,207.67) ;
\draw [shift={(395.17,207.67)}, rotate = 180] [color={rgb, 255:red, 0; green, 0; blue, 0 }  ][fill={rgb, 255:red, 0; green, 0; blue, 0 }  ][line width=0.75]      (0, 0) circle [x radius= 3.35, y radius= 3.35]   ;
\draw [shift={(505,207.67)}, rotate = 180] [color={rgb, 255:red, 0; green, 0; blue, 0 }  ][fill={rgb, 255:red, 0; green, 0; blue, 0 }  ][line width=0.75]      (0, 0) circle [x radius= 3.35, y radius= 3.35]   ;
\draw   (450.26,125.8) -- (505.79,44.6) -- (546.49,146.62) -- cycle ;
\draw    (450.26,125.8) -- (505.79,44.6) ;
\draw [shift={(505.79,44.6)}, rotate = 304.37] [color={rgb, 255:red, 0; green, 0; blue, 0 }  ][fill={rgb, 255:red, 0; green, 0; blue, 0 }  ][line width=0.75]      (0, 0) circle [x radius= 3.35, y radius= 3.35]   ;
\draw [shift={(450.26,125.8)}, rotate = 304.37] [color={rgb, 255:red, 0; green, 0; blue, 0 }  ][fill={rgb, 255:red, 0; green, 0; blue, 0 }  ][line width=0.75]      (0, 0) circle [x radius= 3.35, y radius= 3.35]   ;
\draw    (505.79,44.6) -- (546.49,146.62) ;
\draw [shift={(546.49,146.62)}, rotate = 68.25] [color={rgb, 255:red, 0; green, 0; blue, 0 }  ][fill={rgb, 255:red, 0; green, 0; blue, 0 }  ][line width=0.75]      (0, 0) circle [x radius= 3.35, y radius= 3.35]   ;
\draw [shift={(505.79,44.6)}, rotate = 68.25] [color={rgb, 255:red, 0; green, 0; blue, 0 }  ][fill={rgb, 255:red, 0; green, 0; blue, 0 }  ][line width=0.75]      (0, 0) circle [x radius= 3.35, y radius= 3.35]   ;
\draw   (450.44,126.08) -- (352.55,135.69) -- (404.57,38.96) -- cycle ;
\draw    (450.44,126.08) -- (352.55,135.69) ;
\draw [shift={(352.55,135.69)}, rotate = 174.39] [color={rgb, 255:red, 0; green, 0; blue, 0 }  ][fill={rgb, 255:red, 0; green, 0; blue, 0 }  ][line width=0.75]      (0, 0) circle [x radius= 3.35, y radius= 3.35]   ;
\draw [shift={(450.44,126.08)}, rotate = 174.39] [color={rgb, 255:red, 0; green, 0; blue, 0 }  ][fill={rgb, 255:red, 0; green, 0; blue, 0 }  ][line width=0.75]      (0, 0) circle [x radius= 3.35, y radius= 3.35]   ;
\draw    (352.55,135.69) -- (404.57,38.96) ;
\draw [shift={(404.57,38.96)}, rotate = 298.27] [color={rgb, 255:red, 0; green, 0; blue, 0 }  ][fill={rgb, 255:red, 0; green, 0; blue, 0 }  ][line width=0.75]      (0, 0) circle [x radius= 3.35, y radius= 3.35]   ;
\draw [shift={(352.55,135.69)}, rotate = 298.27] [color={rgb, 255:red, 0; green, 0; blue, 0 }  ][fill={rgb, 255:red, 0; green, 0; blue, 0 }  ][line width=0.75]      (0, 0) circle [x radius= 3.35, y radius= 3.35]   ;
\draw [color={rgb, 255:red, 208; green, 2; blue, 27 }  ,draw opacity=1 ]   (352.55,135.69) -- (395.17,207.67) ;
\draw [color={rgb, 255:red, 208; green, 2; blue, 27 }  ,draw opacity=1 ]   (546.49,146.62) -- (505,207.67) ;
\draw [color={rgb, 255:red, 208; green, 2; blue, 27 }  ,draw opacity=1 ]   (404.57,38.96) -- (505.79,44.6) ;

\draw (33,65.4) node [anchor=north west][inner sep=0.75pt]    {$G:$};
\draw (132,29.4) node [anchor=north west][inner sep=0.75pt]    {$v_{1}$};
\draw (207,132.4) node [anchor=north west][inner sep=0.75pt]    {$v_{2}$};
\draw (53,132.4) node [anchor=north west][inner sep=0.75pt]    {$v_{3}$};
\draw (507,211.07) node [anchor=north west][inner sep=0.75pt]    {$y_{2}$};
\draw (364,199.07) node [anchor=north west][inner sep=0.75pt]    {$y_{3}$};
\draw (518,28.4) node [anchor=north west][inner sep=0.75pt]    {$x_{2}$};
\draw (554,140.4) node [anchor=north west][inner sep=0.75pt]    {$x_{3}$};
\draw (378,18.4) node [anchor=north west][inner sep=0.75pt]    {$z_{3}$};
\draw (328,128.4) node [anchor=north west][inner sep=0.75pt]    {$z_{2}$};
\draw (448,98.4) node [anchor=north west][inner sep=0.75pt]    {$v$};
\draw (178,86.4) node [anchor=north west][inner sep=0.75pt]    {$1$};
\draw (135,153.4) node [anchor=north west][inner sep=0.75pt]    {$2$};
\draw (90,84.4) node [anchor=north west][inner sep=0.75pt]    {$2$};
\draw (295,68.4) node [anchor=north west][inner sep=0.75pt]    {$G'$};
\draw (428,62.4) node [anchor=north west][inner sep=0.75pt]  [font=\scriptsize]  {$2$};
\draw (470,65.4) node [anchor=north west][inner sep=0.75pt]  [font=\scriptsize]  {$1$};
\draw (482,154.4) node [anchor=north west][inner sep=0.75pt]  [font=\scriptsize]  {$1$};
\draw (389,114.4) node [anchor=north west][inner sep=0.75pt]  [font=\scriptsize]  {$1$};
\draw (408,157.4) node [anchor=north west][inner sep=0.75pt]  [font=\scriptsize]  {$2$};
\draw (534,88.4) node [anchor=north west][inner sep=0.75pt]  [font=\scriptsize]  {$2$};
\draw (362,78.4) node [anchor=north west][inner sep=0.75pt]  [font=\scriptsize]  {$2$};
\draw (444,217.4) node [anchor=north west][inner sep=0.75pt]  [font=\scriptsize]  {$2$};
\draw (494,121.4) node [anchor=north west][inner sep=0.75pt]  [font=\scriptsize]  {$2$};
\draw (358,168.4) node [anchor=north west][inner sep=0.75pt]  [font=\scriptsize]  {$3$};
\draw (527.74,180.54) node [anchor=north west][inner sep=0.75pt]  [font=\scriptsize]  {$3$};
\draw (451,25.4) node [anchor=north west][inner sep=0.75pt]  [font=\scriptsize]  {$3$};

\end{tikzpicture}
    \caption{Small example: $G'$ from $G$ with $c = 3$.}
    \label{G'}
\end{figure}

We construct a complete graph $G'$ with edge-costs in $\{1,2,3,4\}$ 
(called a wheel graph in \cite{uttpnp}) as follows:
Fix a positive integer $c$ and choose a vertex $v$ of $G$ to be the 
\textit{central vertex}.
Form the graph $G^{(c)}$ with $c$ copies of $G$ glued together at $v$.
Then set $G'$ and its associated edge-costs to be the metric completion of $G^{(c)}$.
Hence, the cost of an edge between vertices in the same
copy remains the same as it was in $G$, while the cost of an edge $e = (x, y)$
between vertices in different copies is the cost of a minimum-cost
path between $x$ and $y$ given by $D(x,v) + D(v,y)$. 
Figure~\ref{G'} contains an example of the construction where 
$v = v_1$ is the central vertex and $c = 3$. Not all edges of
$G'$ are shown but the full distance matrix $D$ is given.

Note that $G'$ has $m := c(n-1)+1$ vertices. 
For the remainder of the paper, we fix $c \geq 5$.

\begin{align*}
D = 
\kbordermatrix{& v &  x_2 & x_3 & y_2 & y_3 & z_2 & z_3\\
v & - & 1 & 2 & 1 & 2 & 1 & 2\\
x_2 & 1 & - & 2 & 2 & 3 & 2 & 3 \\
x_3 & 2 & 2 & - & 3 & 4 & 3 & 4 \\
y_2 & 1 & 2 & 3 & - & 2 & 2 & 3 \\
y_3 & 2 & 3 & 4 & 2 & - & 3 & 4 \\
z_2& 1 & 2 & 3 & 2 & 3 & - & 2 \\
z_3& 2 & 3 & 4 & 3 & 4 & 2 & -
}
\end{align*}

The next lemma is Lemma 3.3 in \cite{uttpnp}.
\begin{lemma}\label{lemma:cost}
There exists a TSP tour in $G$ with cost at most $K$
if and only if there exists a TSP tour in $G'$ with cost at most $cK$.
\end{lemma}

\begin{cor}\label{cor:cost}
The optimal cost of a TSP tour in $G$
is $K$ if and only if the optimal cost of a TSP tour in $G'$ is $cK$.
\end{cor}

Our reduction from $(1,2)$-TSP to TTP closely follows the one described
in \cite{uttpnp}.  The main difference is in the definition of the edge-costs.

Given an instance of $(1,2)$-TSP on graph $G$, construct $G'$
and the associated edge-costs
as described above. Then construct a new
graph $H$ by adding a new vertex $u$ and joining $u$ to all vertices of
$G'$ with an edge-cost of $w_u$ to be specified depending on the context.
Construct the corresponding TTP instance on $H$ with $10m(m+1)$ teams
by placing $2$ teams at the
central vertex $v$, one team at each of the other $m-1$ vertices of $G'$,
and $(m+1)(10m-1)$ teams at $u$.

Bhattacharyya~\cite{uttpnp} describes a construction of a TTP schedule for 
the TTP instance on $H$ 
from a TSP tour $\tau$ in $G$ which is essentially the following:

First, the teams are split into $10m$ groups $g_1,\ldots,g_{10m}$,
each of size $m+1$, so that every team at a vertex in $G'$ is in $g_1$
and the remaining teams, which are all at $u$, are
arbitrarily divided into $10m-1$ groups of size $m+1$.
We label the teams in $g_1$ as
$t_{1,1},\ldots, t_{1,m+1}$ so that they follow the order in which
they appear in the tour $\tau'$.
For $i = 2,\ldots,10m$, we label
the teams in $g_i$ as $t_{i,1},\ldots, t_{i,m+1}$ arbitrarily.

We build the TTP schedule in two phases.
The first phase consists of $2m$ rounds made up of double round-robin schedules,
one for each group, so that every team plays each of the other $m$ teams in its 
group once at home and once away for a total of $2m$ games.
The double round-robin schedule for each group can be constructed by
using a standard construction for a single-round robin schedule and then
appending to it a copy of itself but with home-away assignments reversed.

In the second phase, every team plays $2(m+1)(10m-1)$ games against teams from
different groups. 
Phase 2 of the schedule is constructed in two stages.
  For the first stage, we treat each of the groups $g_1,\ldots,g_{10m}$
  as a single ``dummy team'' and create a single round-robin ``dummy tournament'' 
  so that $g_1$ plays all away games against $g_2,\ldots,g_{10m}$ in that order.

  A game between dummy teams $g_i$ and $g_j$, where $i > j$, ``induces''
  the following actual games:
  $t_{i,k}$ plays against 
  $t_{j,k},\ldots,t_{j,m+1},t_{j,1},\ldots,t_{j,k-1}$ in that order.
  (Therefore,
  $t_{j,k}$ plays against
  $t_{i,k},t_{i,k-1},\ldots,t_{i,1},t_{i,m+1},t_{i,m},\ldots,t_{j,k+1}$
  in that order.)
  The home-away assignments of the actual games are inherited from
  the home-away assignments of the dummy teams. In other words,
  if dummy team $g_i$ is at home and receives dummy team
  $g_j$, then all teams in $g_i$ are at home for their games against
  teams in $g_j$. 

  For the second stage, we duplicate the single round-robin dummy tournament of
  the first stage with the home-away assignments reversed. 

  Note that the constructed schedule satisfies the no-repeaters constraint.

\begin{lemma}\label{lemma:ubound}
If the cost of the TSP tour $\tau$ is at most $K$,
then the constructed TTP tournament
has cost at most $20w_u m(m+1)+(10m-1)mcK+8m(m+1).$
\end{lemma}

\begin{proof}
By Lemma~\ref{lemma:cost}, $G'$ has a tour $\tau'$ of cost $\nu \leq cK$. 

We first bound the travel cost of Phase 1.
  Note that games between teams in groups other than $g_1$
  carry a traveling cost of zero since all these teams are at $u$.
  For a team in group $g_1$, the maximum distance
  it travels in these rounds is $8m$ since an edge in $G'$ has cost at
  most $4$, and it must visit $m$ teams. Therefore the total distance travelled
  by all teams in Phase 1 is at most $8m(m+1)$. 

We now bound the travel cost of Phase 2.
The first stage of this phase has $10m-1$ rounds.
  Since there are $m+1$ teams in each group, teams in $g_1$ play a total of
  $(10m-1)(m+1)$ away games at $u$ in this stage.
  They visit $u$, collectively incurring a traveling cost of $(m+1)w_u$ since the
  cost of traveling from any vertex in $G'$ to $u$ is $w_u$.
  Once they are at $u$, they play all teams in $g_2$, then those in $g_3$,
  and so on,
  while remaining at $u$. When they return home, they again
  collectively travel a distance of $(m+1)w_u$. All other groups are at $u$ so no
  distance is travelled in this stage. It follows that the total
  distance travelled in this stage is $2(m+1)w_u$. 

In the second stage, games
  between teams in groups other than $g_1$ incur no travel cost as before.
  But now, all the teams
  in groups $g_2,\ldots,g_{10m}$ must travel
  from $u$ to $G'$ to play teams in $g_1$.
For each $j \in \{2,\ldots, 10m\}$, the games
  between the teams in $g_j$ and those in $g_1$ are, by construction,
  as follows:
  For each $k \in \{1,\ldots,m+1\}$, team $t_{j,k}$ plays the teams
  $t_{1,k},\ldots,t_{1,m+1},t_{1,1},\ldots,t_{1,k-1}$ in that order.
  Hence, team $t_{j,k}$ travels from $u$ to $G'$, then travels along
  the tour $\tau'$ but missing the edge between $t_{1,k-1}$ and $t_{1,k}$,
  and then returns home to $u$.
  As a result, the total travel cost incurred by teams in $g_j$ is 
  $(m+1)(2w_u + \nu) - \nu = 2(m+1)w_u + m\nu.$

In sum, the constructed TTP schedule has cost at most 
\begin{align*}
      &(10m - 1)(2w_u (m+1) + m\nu) + 2w_u (m+1) + 8m(m+1)\\
   \leq\, &(10m - 1)(2w_u (m+1) + mcK) + 2w_u (m+1) + 8m(m+1)\\
   =\,& 20w_um(m+1) + (10m-1)mcK + 8m(m+1)
\end{align*}
as desired.

\end{proof}

\begin{lemma}\label{lemma:construct}
  Given a schedule for the TTP instance $H$ of cost $\nu'$,
  a tour in $G$ having cost $K$ satisfying
  $\nu' \geq 10m(m+1)(2w_u + cK - 4)$
  can be obtained in polynomial time.
\end{lemma}
\begin{proof}
Consider a schedule for the TTP instance $H$ of cost $\nu'$.

Observe that the travel pattern of a team $t$ induces
  a closed walk $P_t$ through all the vertices of $H$
  having cost equal to the total travel cost of team $t$.
  Choose $t$ so that this cost, which we call $M$, is least possible.
  Clearly, $t$ can be obtained in polynomial time.

  As there are $10m(m+1)$ teams in total and each team's travel cost
  is at least $M$, we have $\nu' \geq 10m(m+1)M$.
  To complete the proof, we construct from $P_t$ a tour in $G$ of cost 
  $K$ satisfying $2w_u + cK - 4 \leq M.$

  First of all, if $P_t$ is not a tour in $H$, there must be edges $(x,z)$ 
  and $(z,y)$ so that $x,z,y$ appear in sequence in the walk and $z$ is visited
  more than once by the walk. As the edge-costs form a metric, we modify the walk
  without increasing its cost
  by replacing $(x,z)$ and $(z,y)$ with $(x,y)$ so that the walk now skips
  $z$ and goes directly from $x$ to $y$. We repeat this short-cutting
  process until the walk
  is a tour in $H$. Remove the edges in this tour incident with $u$ and join
  the other ends to obtain a tour $T'$ in $G'$.
  Since the cost of an edge in $G'$ is at most 4,
  the cost of this tour is at most $M-2w_u+4$.

  By our choice of the edge-costs for $G'$,
  the cost of going from a vertex $x$ in one copy of $G$ in $G^{(c)}$ to 
  a vertex $y$ in another copy is the same as the cost of
  going from $x$ to $v$ and then from $v$ to $y$.
  Hence, we can turn the tour $T'$ into to a closed walk $P'$ by replacing edges 
  $(x,y)$ that cross copies by the edges $(x,v)$, $(v,y)$
  while maintaining the same cost.

  Observe that in each copy of $G$, $P'$ induces a collection of tours that
  span all the vertices in that copy of $G$ meeting only at $v$.
  As the edge-costs form a metric, we can again perform short-cutting
  without increasing the cost of $P'$
  until $P'$ induces a tour in every copy of $G$.
  We pick the induced tour with minimum cost and call its cost $K$.
  Clearly, this tour can be obtained in polynomial time.

  Since there are $c$ copies of $G$ in $G'$,
  the cost of $P'$ is at least $cK$.
  But we know that the cost of $P'$ is at most $M-2w_u+4$.
  Hence, $M \geq 2w_u +cK - 4$ as desired.
\end{proof}

\section{APX completeness of UTTP}

The goal of this section is to establish the following:

\begin{thm}\label{thm:main}
  The unconstrained Traveling Tournament Problem, with or without the no-repeaters
  constraint, is APX-complete.
\end{thm}

The core of our proof is an L-reduction from a boosted version of
$(1,2)$-TSP defined as follows:
Let $G$ be a complete graph with $m$ nodes with edge-costs in $\{1,2\}$.  
Let \textit{boosted $(1,2)$-TSP} be the problem of computing a TSP
tour $T$ such that the objective $(m(m+1))(\text{cost}(T))$ is minimized.

\begin{lemma}
  Boosted $(1,2)$-TSP is APX-complete.
\end{lemma}
\begin{proof}
 Let $\mu_B$ and $\mu$ be the optimal
  values of the boosted $(1,2)$-TSP and regular $(1,2)$-TSP, respectively, 
  on a given complete graph $G$ with $m$ nodes with edge-costs in $\{1,2\}$.  

  Suppose there is a polynomial-time approximation scheme for boosted $(1,2)$-TSP,
  then for any $\varepsilon$ we can find a solution $S$ in polynomial time such
  that $m(m+1)\cdot \text{cost}(S)$ is at most $(1+ \varepsilon)\cdot \mu_B =
  (1+\varepsilon)(m(m+1))\cdot \mu$. Therefore, $\text{cost}(S)$ is at most
  $(1+\varepsilon)\cdot \mu$. Moreover, since $S$ is a TSP tour on $G$, it is a
  solution to the regular TSP instance. It follows that if boosted $(1,2$)-TSP
  has a PTAS, then for every $\varepsilon$, we can find a TSP tour $S$ such
  that $\text{cost}(S)$ is within $(1+\varepsilon)$ of $\mu$. Since $(1,2)$-TSP 
  is APX-complete by Proposition~\ref{prop:PY}, we conclude that 
  boosted $(1,2)$-TSP is APX-complete as well.
\end{proof}

Let $I$ be a boosted TSP instance on graph $G$ with input integer $c\geq 5$.
Let $I'$ be the corresponding TTP instance constructed as in Section
\ref{construction}.

\subsection{Proof of Theorem~\ref{thm:main}}
First, note that UTTP, with or without the no-repeaters constraint,
admits an $O(1)$-approximation algorithm~\cite{uttp275} and is therefore in APX.

To complete the proof, we show that
there is an L-reduction from
boosted $(1,2)$-TSP to UTTP
by making use of the results of Section \ref{construction}.
In particular, we verify the following conditions:
\begin{enumerate}
  \item For each instance $I$ of boosted TSP, we can compute an instance $I'$
    of TTP in polynomial time.
  \item $\opt(I') \leq 40c \cdot \opt(I)$
  \item Given a solution of value $\nu'$ to $I'$, we can compute in polynomial
    time a solution to $I$ of cost $\mu$ such that \begin{equation*}
      \mu - \opt(I) \leq \nu'-\opt(I').
  \end{equation*}
\end{enumerate}

To satisfy condition $1$, it suffices to use the construction in Section
\ref{construction} where the boosted $(1,2)$-TSP instance is the TSP instance
on $G$ with its objective multiplied by $m(m+1)$.

For condition $2$, suppose that the instance of boosted
$(1,2)$-TSP on the graph $G$ has optimal value $m(m+1)\mu^*$.
Hence, the optimal value of the regular TSP instance is $\mu^*$.
Clearly, $n \leq \mu^* \leq 2n$.
Applying Lemma~\ref{lemma:ubound} with $w_u = \frac{2cn - 1}{2}$ gives
\begin{align*}
  \opt(I') &\leq 20w_u m(m+1) + (10m-1)mc\mu^* + 8m(m+1) \\
    &= 10(2cn-1) m(m+1) + (10m-1)m(2cn) + 8m(m+1)\\
    &< 40m(m+1)cn \\
    &\leq 40m(m+1)c\mu^*\\
    &= 40c \cdot \opt(I).
\end{align*}

Finally, we verify condition 3.
Consider a schedule for the TTP instance with cost $\nu'$.
By Lemma~\ref{lemma:construct},
we can construct, in polynomial time,
a tour in $G$ having cost $\mu'$ satisfying
\begin{equation*}
\nu' \geq 10m(m+1)(2w_u + c\mu' - 4).
\end{equation*}

Let $\mu = m(m+1)\mu'$.
If $\mu = \opt(I)$, then there is nothing to prove.
Hence, assume that $\mu > \opt(I)$.

Let $\opt(I)$ be given by $m(m+1)\mu^*$.
Observe that $\mu^*$ is the optimal value of the regular TSP instance on $G$.
Thus, $\mu' > \mu^* \geq n \geq 3$.
By Lemma~\ref{lemma:ubound}, there is a TTP schedule having cost
\begin{equation*}
\nu^* \leq 20w_u m(m+1) + (10m-1)mc\mu^* + 8m(m+1).
\end{equation*}
Then,
\begin{align*}
  \nu' - \opt(I')
  & \geq \nu' - \nu^* \\
  & \geq 10m(m+1)(2w_u + c\mu' - 4) \\
  & \quad - \left(20w_u m(m+1)+(10m-1)mc\mu^*+8m(m+1)\right) \\
  & = 10m(m+1)(c\mu' - 4) - (10m-1)mc\mu^* - 8m(m+1) \\
  & = (10m-1)m(c(\mu'-\mu^*) - 4) + 11m(c\mu' - 4) - 8m(m+1) \\
  & > (10m-1)m(c(\mu'-\mu^*) - 4) + 11m(cn - 4) - 8m(c(n-1)+2) \\
  & = (10m-1)m(c(\mu'-\mu^*) - 4) + m(c(3n+8)-60) \\
  & \geq (10m-1)m(c-4)(\mu'-\mu^*)
\end{align*}
since $c \geq 5$,, $\mu' > \mu^*$ and $\mu'$ and $\mu^*$ are integers.
Suppose by way of contradiction that
$\mu - \opt(I) > \nu' - \opt(I').$
In other words, suppose that
\begin{equation*}
  m(m+1)(\mu' - \mu^*) > \nu' - \opt(I').
\end{equation*}
Continuing the above derivations gives
\begin{align*}
  \nu' - \opt(I')
	& > (10m-1)m(c-4)\left(\frac{\nu' - \opt(I')}{m(m+1)}\right) \\
  & = \frac{(10m-1)(c-4)}{m+1}(\nu' - \opt(I')) \\
  & > \nu' - \opt(I'),
\end{align*}
a contradiction.


\bibliographystyle{elsarticle-num-names} 
\bibliography{cas-refs}





\end{document}